\documentclass[useAMS,usenatbib]{mn2e}
\usepackage{graphicx,aas_macros,ulem} 
\usepackage{color} 

\input epsf

\renewcommand{\em}{\it}  

\newcommand{\sfig}[2]{\includegraphics[width=#2]{#1}}
\newcommand{\Sfig}[2]{
    \begin{figure}
    \sfig{#1.eps}{\columnwidth}
    \caption{{\small #2}}
    \label{fig:#1}
    \end{figure}
}
\newcommand{\Sfigtwo}[3]{
    \begin{figure}
    \sfig{#1.eps}{\columnwidth}
    \sfig{#2.eps}{\columnwidth}
    \caption{{\small #3}}
    \label{fig:#1}
    \end{figure}
}

\def\LCDM{$\Lambda$CDM~}

\newcommand{\ngal}{n_{\rm gal}}

\newcommand{\zmed}{z^{\rm med}}

\newcommand{\rmsshear}{\sigma_\gamma}
\newcommand{\rmsflexF}{\sigma_\flexF}
\newcommand{\rmsflexG}{\sigma_\flexG}

\newcommand{\FTkappa}{\tilde\kappa}
\newcommand{\FTshear}{\tilde\gamma}
\newcommand{\FTflexF}{\tilde\flexF}
\newcommand{\FTflexG}{\tilde\flexG}

\usepackage{bm}

\newcommand{\dx}[1]{{\rm d} #1 \,}
\newcommand{\dd}[1]{d #1\,}

\def\vl{\bm l} 

\newcommand{\flexF}{\mathcal{F}}
\newcommand{\flexG}{\mathcal{G}}
\newcommand{\BHB}{{\rm BHB}}
\newcommand{\kappaBHB}{\kappa_{\rm BHB}}
\newcommand{\kappasignal}{\kappa_{\theta}}

\newcommand{\kappaMS}{\kappa_\Theta}
\newcommand{\kappamap}{\kappa_{\rm map}}
\newcommand{\kappadeep}{\kappa_{\rm deep}}
\newcommand{\kappawide}{\kappa_{\rm wide}}
\newcommand{\skypos}{\mathbf{x}}
\newcommand{\sigflex}{\sigma_{\flexF\flexG}}
\newcommand{\smoothangle}{\theta}
\newcommand{\fieldsize}{\Theta}
\newcommand{\noise}{\nu}
\newcommand{\pgal}{p_{\rm gal}}
\newcommand{\kkern}{\tilde\mathcal{D}}
\newcommand{\remain}{\mathcal{R}}

\newcommand{\mean}[1]{\left\langle{#1}\right\rangle}

\newcommand{\beq}{\begin{equation}}
\newcommand{\eeq}{\end{equation}}
\newcommand{\eeqp}{\;.\end{equation}}
\newcommand{\eeqc}{\;,\end{equation}}
\newcommand{\bea}{\begin{eqnarray}}
\newcommand{\eea}{\end{eqnarray}}

\newcommand{\refeq}[1]{(\ref{eq:#1})}

\newcommand{\reffig}[1]{Figure \ref{fig:#1}}

\title[Delensing Sirens]
{Delensing Gravitational Wave Standard Sirens with Shear and Flexion Maps }
\author[C. Shapiro et al.\ ]{
C. Shapiro$^{1}$\thanks{charles.shapiro@port.ac.uk}\footnotemark[0],
D. J. Bacon$^{1}$,
M. Hendry$^{2}$,
and B. Hoyle$^{1}$ \\
$^{1}$Institute of Cosmology \& Gravitation, University of Portsmouth, Dennis Sciama Bldg., Portsmouth, PO1 3FX, UK\\
$^{2}$Department of Physics \& Astronomy, University of Glasgow, Kelvin Bldg., Glasgow,  G12 8QQ, UK
}
\begin{document}
\date{Version as of \today}

\pagerange{\pageref{firstpage}--\pageref{lastpage}} \pubyear{2009}

\maketitle

\label{firstpage}

\begin{abstract}
Supermassive black hole binary systems (SMBHB) are standard sirens -- the gravitational wave analogue of standard candles -- and if discovered by gravitational wave detectors, they could be used as precise distance indicators.  Unfortunately, gravitational lensing will randomly magnify SMBHB signals, seriously degrading any distance measurements. Using a weak lensing map of the SMBHB line of sight, we can estimate its magnification and thereby remove some uncertainty in its distance, a procedure we call ``delensing.''  We find that delensing is significantly improved when galaxy shears are combined with flexion measurements, which reduce small-scale noise in reconstructed magnification maps.  Under a Gaussian approximation, we estimate that delensing with a 2D mosaic image from an Extremely Large Telescope (ELT) could reduce distance errors by about 30-40\% for a SMBHB at $z=2$.  Including an additional wide shear map from a space survey telescope could reduce distance errors by 50\%.  Such improvement would make SMBHBs considerably more valuable as cosmological distance probes or as a fully independent check on existing probes.
\end{abstract}


\section{Introduction}

Recently there has been growing interest in the potential future use
of coalescing compact binary systems as high-precision cosmological
distance indicators. During their inspiral phase, prior to coalescence,
the amplitude and frequency of the gravitational wave emission from these
systems varies rapidly with time, following a characteristic `chirp'
waveform that is strongly dependent on the binary masses.

In a seminal paper, Schutz (\citeyear{schutz_1986}) showed how measurement of the
amplitude, frequency and frequency derivative of the inspiralling binary,
using observations carried out by a network of interferometric
gravitational wave detectors, could yield a precise estimate of its
luminosity distance -- completely independent of the traditional cosmic
distance ladder. More recently these binary systems have been labelled
`gravitational wave standard sirens' -- by analogy with electromagnetic
standard candles (although, strictly, they do {\em not\/} require the
assumption that all sources have the same gravitational wave luminosity).

Siren candidates include neutron star--neutron star binaries, believed to
be associated with short-duration gamma ray bursts \citep{eichler_livio_etal_1989}. 
These systems are among the prime targets for detection by the next 
generation of ground-based interferometers, Advanced LIGO and Advanced
VIRGO, which should detect up to a few dozen such binary coalescences per 
year \citep{kopparapu_hanna_etal_2008}. Recently \citet{nissanke_hughes_etal_2009}, extending 
earlier work by \citet{dalal_holz_etal_2006}, have shown that the advanced detectors 
will be able to determine the luminosity distance of these sirens to an 
accuracy of better than 30\%, out to a distance of 600 Mpc.

Even more promising siren candidates are the mergers of supermassive black 
hole binaries (SMBHBs).  These systems are expected to be extremely
luminous gravitational wave sources and will be prime observational
targets for the planned {\em Laser Interferometer Space Antenna\/} 
 \citep[{\em LISA\/};][]{bender_etal_1994}. Moreover, the issues addressed in this paper are of 
particular importance for SMBHBs; consequently they will be the principal 
focus in what follows.

The expected number and redshift distribution of SMBHBs that will be
observed by LISA is very uncertain, being strongly dependent on the details
of our model for the history of galaxy mergers \citep{sesana_volonteri_etal_2007, arun_babak_etal_2009}. However, the impact of observing even a 
handful of SMBHBs could be very significant, as first demonstrated by 
\citet[hereafter HH05]{holz_hughes_2005}. Those authors performed a Monte Carlo 
study to calculate the fractional accuracy with which luminosity distance 
can be determined for sirens of a range of masses and at different reshifts, 
randomly distributed in orientiation and sky position. Assuming that the sky 
position of the SMBHB could be determined exactly from electromagnetic 
observations, HH05 found that, for e.g. the merger of a $10^5$ solar mass 
and $6 \times 10^5$ solar mass black hole at $z=1$, the most probable
fractional error on luminosity distance was $~ 0.1\%$.  Moreover, the same
binary system observed at $z=3$ would yield a most probable fractional error 
of $~0.5\%$.  HH05 then demonstrated the dramatic cosmological potential of 
these high-precision distance indicators: only two sirens observed at $z=1$ 
and $z=3$ could constrain cosmological parameters at a level competitive with 3000 type Ia supernovae.

As pointed out by HH05, however, there is a huge caveat: standard sirens
(like standard candles) will be \hbox{(de-)magnified} by weak lensing caused by matter fluctuations along their line of sight.  The observed luminosity distance to a siren is related to its true luminosity distance by
\beq \label{eq:magdist}
D_L^{\rm obs} = D_L\, \mu^{-1/2} \approx D_L(1-\delta\mu/2)
\eeq
where $\mu$ is the lensing magnification and $\delta\mu\equiv\mu-1=0$ for an unlensed source.  The scatter in $\delta\mu$ is expected to be several percent for high-redshift sirens; this adds in quadrature to the sirens'
intrinsic scatter, substantially degrading their effectiveness as cosmological
probes.
Moreover, as shown by \citet{kocsis_frei_etal_2006}, weak lensing will also
significantly increase the extent in redshift of the 3-D error box 
determined by LISA -- thus rendering much more difficult the task of 
identifying an electromagnetic counterpart.

Subsequent work on SMBHBs has mainly focussed on better localising and 
understanding their electromagnetic signatures \citep[see e.g.][]{kocsis_loeb_2008, lang_hughes_2006} and further reducing their intrinsic scatter in 
the absence of lensing. \citet{lang_hughes_2006} included spin-induced precession 
\citep{vecchio_2004} in their model of the siren waveform and showed that this 
significantly reduced the median luminosity distance error, and angular size 
of the LISA error box, because the precession terms break the degeneracies 
between other waveform parameters. Other authors \citep[see e.g.][]{porter_cornish_2008} have found similar improvements through the inclusion of higher-order 
harmonics in the siren waveform, and recently \citet{arun_babak_etal_2009} have 
considered {\em both\/} precession and higher harmonics in their modelling 
of SMBHBs, with impressive results.

Notwithstanding this progress, however, the damaging impact of weak lensing
on the performance of sirens has remained largely unaddressed -- although 
a notable exception has been \citet{jonsson_goobar_etal_2007},
which built on earlier work exploring weak lensing corrections 
to standard candles \citep{jonsson_dahlen_etal_2005, jonsson_dahlen_etal_2006a, jonsson_dahlen_etal_2006}.  Their method proposes to model, and thus correct for, 
lensing magnification by using
the observed photometric and spectroscopic properties of foreground galaxies 
along the line of sight to the siren. The authors consider the case where the lensing signal 
is dominated by a single dark matter halo; the magnification factor is 
computed from an inferred halo mass profile, which is in turn constrained by  photometric and spectroscopic data using the Tully-Fisher or Faber-Jackson 
relations. The authors' results are quite impressive, showing
that the dispersion due to lensing for a standard candle or siren at $z=1.5$
can be reduced by about a factor of two \citep{jonsson_mortsell_etal_2009}. However, their method does involve a number of specific modelling assumptions about halo mass profiles and the observable mass proxy.

In this paper we pursue an alternative approach which makes fewer modeling assumptions: we propose calculating weak lensing corrections derived from gravitational lensing shear and flexion maps constructed in the direction of the SMBHB.
In other words, we directly measure the siren magnification in order to remove it, a procedure we call ``delensing''.
Dalal et al.\  (\citeyear[][hereafter D03]{dalal_holz_etal_2003}) have shown that performing such corrections for large surveys of type Ia supernovae (SNeIa) is unlikely to be worthwhile.  Although SNeIa surveys conveniently provide galaxy shape measurements that could be used to delens each supernova, they will not have galaxy densities high enough to resolve small-scale contributions to the magnification.

We find at least three reasons to reconsider delensing in the case of SMBHBs:
\begin{itemize} 
\item Since we expect SMBHB detections to be rare, we can furnish each one with a very deep pointed follow-up observation, thus obtaining the high galaxy densities needed for high-resolution shape maps.
\item Galaxy images from devoted observations could enable higher quality shear and flexion measurements than those available from a survey telescope.  We will show that high-quality flexion measurements are a substantial advantage.
\item  Delensing will make more of an impact on SMBHB distance measurements than on SNeIa measurements.  Lensing will be by far the largest source of scatter in SMBHB distances, whereas SNeIa retain a significant intrinsic scatter even after delensing.  In addition, since surveys will detect a very large number of SNeIa, the magnification noise will be partially averaged away.
\end{itemize}

In this paper we investigate the efficacy of delensing SMBHBs with high-quality shear and flexion maps.  In \S\ref{mapmaking}, we briefly review the Kaiser-Squires technique for reconstructing a magnification map from shape measurements.  In \S\ref{methodology}, we present our methodology for estimating how precisely a SMBHB magnification can be measured from a set of shape maps.  We also demonstrate the advantage of combining shear and flexion measurements to substantially reduce shape noise.  In \S\ref{results}, we compute the reduction in the scatter in SMBHB distances that could be achieved by delensing with an Extremely Large Telescope (ELT).  We further show that delensing can be improved by combining ELT maps with a wide space telescope survey.  We conclude in \S\ref{conclusions} along with a discussion of topics for further investigation.

\section{Mapmaking} \label{mapmaking}

In order to improve our estimate of the luminosity distance to the siren, $D_L$, we need to obtain an accurate estimate of the lensing magnification $\mu(\skypos)$ at the siren's position $\skypos$ on the sky. If the siren signal has undergone weak lensing, this magnification is related to the lensing convergence $\kappa$ by $\mu \simeq 1+2\kappa$. So if we can make an accurate convergence map at and around position $\skypos$, we have an estimate of the magnification in that region.

Convergence maps can be constructed from shear measurements using the Kaiser-Squires (KS) inversion method \citep{kaiser_squires_1993,bartelmann_schneider_2001}. Since shear and convergence are both combinations of derivatives of the lensing potential, the convergence can be estimated from the shear by averaging shear estimators in pixels on a grid, Fourier transforming to obtain shear estimators $\FTshear_i(\vl)$ and finding Fourier-transformed convergence $\FTkappa$ according to

\begin{equation}
\FTkappa({\vl})=\frac{l_1^2-l_2^2}{l^2}\tilde\gamma_1+2 \frac{l_1 l_2}{l^2}\tilde\gamma_2
\end{equation}
This can then be inverse Fourier transformed to obtain a $\kappa$ map, and hence $\mu$ map, including the value at $\skypos$. 

Higher order shape distortions can similarly be inverted to create a $\kappa$ map. Bacon et al.\  (\citeyear{bacon_goldberg_etal_2006}) showed that flexion measurements, characterizing the slight arc of images, can be inverted in a fashion similar to Kaiser-Squires, with the prescription

\begin{equation}
\FTkappa(\vl)=\frac{il_1}{l^2}\tilde \flexF_1+\frac{il_2}{l^2}\tilde \flexF_2
\end{equation}
\begin{equation}
\FTkappa(\vl)=i\frac{l_1^3-3l_1 l_2^2}{l^4}\tilde \flexG_1+i\frac{l_2^3-3l_1^2 l_2}{l^4}\tilde \flexG_2
\end{equation}
where $\flexF$ is the 1-flexion, representing a skew to the galaxy shape, while $\flexG$ is the 3-flexion, representing the trefoil component of the galaxy.

If we had perfect knowledge of the $\gamma$, $\flexF$ and/or $\flexG$ field on all scales, and at the redshift of the SMBHB, this approach would yield a perfect map for $\mu$ and hence a perfect correction for $D_L$. However in reality our measurements of lensing distortions are noisy, exist only on lines of sight which contain a galaxy, cover a range of redshifts, and are for a limited patch of sky, so our $\mu$ map will be imperfect. 

Firstly there is the impact of noisy shape measurements. We cannot measure the true shear or flexion of a galaxy; rather, we measure an estimator for these quantities, including the noise due to the galaxy's intrinsic shape (ellipticity, skew etc), background noise from the sky and the camera, and noise due to an imperfect deconvolution of the PSF. This means that each galaxy's estimate for shear or flexion has a net error denoted by $\rmsshear$, $\rmsflexF$ and $\rmsflexG$ respectively; this includes the dispersion from all the above effects. Then the error on the signal in a pixel on our shear/flexion map will be $\sigma/\sqrt{\ngal A_{\rm pix}}$, where $\sigma$ is the galaxy-by-galaxy error in the relevant distortion, $\ngal$ is the number density of galaxies and $A_{\rm pix}$ is the area of the pixel. In \S\ref{shapenoise} we will see how this translates to an error on $\tilde \kappa$ in Fourier space. 

We note that the finite number density of galaxies acts as a hard limit on the scales we can probe for $\mu$; if there are density fluctuations on scales smaller than the inter-galaxy spacing, we will not be able to estimate their lensing effect.  As shown by D03, this is a limiting factor when delensing SNeIa since Cold Dark Matter models predict a significant level of sub-arcminute fluctuations in $\mu$.  On a related note, we will have to smooth our maps on small scales to overcome the intrinsic shape noise; again, this will restrict the scales on which we can describe $\mu$.

In addition to these small-scale inaccuracies, a realistic survey will have a finite size on the sky. This means that large-scale density modes, which again contribute to the total $\mu$, will not be estimated by our survey. A related issue is the mass-sheet degeneracy; $\kappa$ estimated from Kaiser-Squires inversion is degenerate with $(1-\beta)\kappa + \beta$ for a constant $\beta$. For large fields $\beta\rightarrow 0$, but small surveys need to take account of this effect. 
Our method for estimating errors in $\mu$ must therefore deal with small scale smoothing and large-scale cut-offs in our data. In addition, the inversion equations estimating $\kappa$ from shear and flexion should be combined in an optimal fashion, as we explain in \S\ref{shapenoise}.

Already it can be seen that ideally we require an ambitious lensing follow-up programme for sirens. This will include observations of the immediate region of the siren, with the best possible resolution (to minimise the $\sigma$s) and depth (to maximize $\ngal$); this will allow us to probe the small-scale contributions to $\mu$.  In addition, we preferably require a large survey (still with considerable image resolution and depth) around the siren position, to probe the large-scale contributions to $\mu$. These requirements could be fulfilled by using an Extremely Large Telescope \citep{elt_science_2006} with adaptive optics to image the central region, together with a survey space telescope to give the large-scale map. We will explore the consequences of this strategy in \S\ref{results_hybrid}.


\section{Methodology} \label{methodology}

\subsection{Estimating reduction in lensing scatter}

Consider a circular field of radius $\fieldsize$ containing galaxies with a redshift distribution $\pgal(z)$.  Suppose we perform a KS-like inversion on shear and flexion maps of the field after smoothing these with a radially symmetric filter with scale-radius $\smoothangle$ (filter can be a top-hat, Gaussian, etc.).  The reconstructed convergence at sky position $\skypos$ is
\beq \label{eq:mapdef}
\kappamap(\skypos) = \kappasignal(\skypos) - K(\fieldsize) + \noise_\smoothangle(\skypos;\fieldsize)
\eeqp
Here, $\kappasignal$ is the true convergence of our galaxies smoothed by our filter, $\noise_\smoothangle$ is filtered shape noise, and $K(\fieldsize)$ is a constant determined by fluctuations larger than the field area.  The mass-sheet degeneracy prevents us from determining this constant from shape measurements alone (see appendix \ref{mass-sheet}).  A simple theoretical estimate of $K$ is the true convergence field at the center of our map, smoothed on the scale of our map:
\beq \label{eq:MSconst}
K(\fieldsize)\approx\kappaMS(\skypos_0)
\eeqp
Note that the noise at $\skypos$ depends on all galaxies in the survey area, not just those within the smoothing scale, because of the non-local properties of the KS inversion.

Let there be a SMBHB with redshift $z_\BHB$ located at $\skypos_\BHB$, and let $\kappaBHB$ be the true convergence along the line of sight.  Our uncertainty in the true convergence, $\sigma(\kappaBHB)$, is initially (i.e. with no lensing map) given by
\beq
\sigma(\kappaBHB) = \mean{\kappaBHB^2}^{1/2}
\eeq
where angular brackets denote an ensemble average.  Once we map the convergence field, we replace our uncertainty with a smaller one, $\sigma^\prime(\kappaBHB)$, as the lensing map allows us to remove some of the true convergence, leaving a residual due to the noise, smoothing, and broad redshift range of the lensing map.  Following D03  and Bower (\citeyear{bower_1991}), the expected improvement is given by
\beq
\sigma^\prime(\kappaBHB)^{2} = (1-r^2)\sigma(\kappaBHB)^2
	\label{eq:rdef}
\eeq
\beq
r^2 \equiv \frac{\mean{\kappamap\kappaBHB}^2}{\mean{\kappaBHB^2}\mean{\kappamap^2}}
	= \frac{\mean{\kappasignal\kappaBHB-\kappaMS\kappaBHB}^2}
	{\mean{\kappaBHB^2}\mean{\kappasignal^2-\kappaMS^2+\noise_\smoothangle(\fieldsize)^2}}
	\label{eq:rdef_terms}
\eeq
where dependence on $\skypos_\BHB$ has been omitted.  Thus $r^2$ can range from zero to unity with $r^2=1$ when our map perfectly reconstructs the convergence at $\skypos_\BHB$.  We are assuming that the lensing fields are Gaussian random; hence we have assumed that fluctuations within the map don't correlate with those larger than the survey area: $\mean{\kappamap\kappaMS}=0$.  We also ignore intrinsic alignments which would correlate lensing with shape noise \citep{crittenden_natarajan_etal_2001, heymans_heavens_2003, hirata_seljak_2004}, and we ignore higher order lensing effects such as reduced shear/flexion \citep{dodelson_shapiro_etal_2006,white_2005,schneider_er_2008}.  These are simplifications that should be revisited in future work.

The various correlations ``$\mean{\kappa\kappa}$'' in \refeq{rdef_terms} can be calculated using the following general expression:
\beq
\mean{\kappa_i\kappa_j} = \frac{1}{2\pi}\int_0^\infty\dx{l}l P_\kappa(l;i,j)\tilde F_i(l)\tilde F_j(l)
\label{eq:KKgeneral}
\eeq
for $i, j \in \{\theta,\fieldsize,\BHB\}$.  The multipole $l$ is the Fourier conjugate to $\skypos$, and we are working with the flat-sky and Limber approximations.  $P_\kappa(l;i,j)$ is a generalized convergence power spectrum, defined below.  $\tilde F_i$ is the Fourier transform of the appropriate filter function for $\kappa_i$: if we assume Gaussian-smoothed shape maps and a top-hat-shaped survey, then
\newcommand{\eqand}{,\hspace{.3cm}}
\beq
\tilde F_\smoothangle(l)=e^{-l^2\theta^2/8} \eqand \tilde F_\fieldsize(l)=\frac{2J_1(l\fieldsize)}{l\fieldsize} \eqand\tilde F_\BHB(l) = 1
\eeqp
The generalized convergence power spectrum is given by an integral over comoving distance $\chi$:
\bea
P_\kappa(l;i,j) &\equiv& \int_0^{\chi_H} \frac{\dd{\chi}}{\chi^2} W_i(\chi)W_j(\chi) P_\delta\left(k;\chi\right) \\
W_i(\chi) &=& \frac{3}{2}\Omega_m H_0^2\frac{\chi}{a(\chi)}\int_{\chi}^{\chi_H} \dd{\chi_s} p_i(z)\frac{dz}{d\chi_s}\frac{\chi_s-\chi}{\chi_s}
\eea
where $P_\delta\left(k;\chi\right)$ is the 3D matter power spectrum for $k=\ell/\chi$ at a distance $\chi$, accounting for the growth of structure.  The scale factor is $a=(1+z)^{-1}$, and $\chi_H$ is the distance to the horizon.  We have assumed a flat Universe for simplicity.  Here, $p_i(z)$ is the redshift distribution of the sources:
\beq
p_\theta(z) = p_\fieldsize(z) = \pgal(z) \eqand p_\BHB = \delta_D(z-z_\BHB)
\label{eq:sources}
\eeqc
where the Dirac delta function, $\delta_D$, should not be confused with the matter overdensity, $\delta$.  Using equations \refeq{KKgeneral}--\refeq{sources}, we can calculate $r^2$ from \refeq{rdef_terms}; we only need $\mean{\noise_\smoothangle(\fieldsize)^2}$, the variance at a point in our convergence map due to noise.





\subsection{Shape Noise in a KS Inversion with Flexion}
\label{shapenoise}

\Sfig{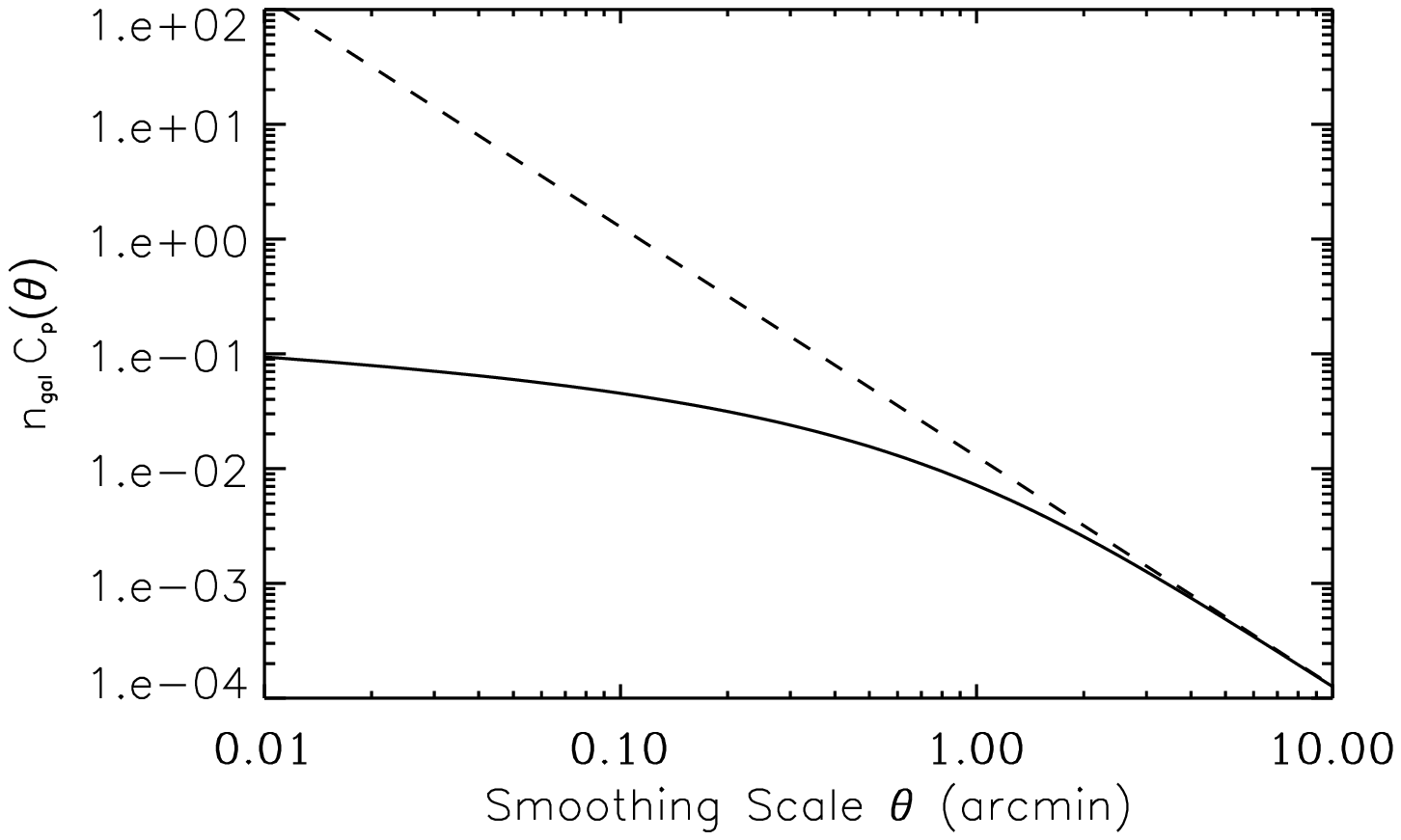}{Variance due to shape noise for a $\kappa$ map reconstructed from shape maps smoothed by a Gaussian filter of width $\smoothangle$.  The dashed line assumes shear only while the solid line adds both types of flexion.  The RMS shapes are those expected for an Extremely Large Telescope: $\rmsshear=0.2, \rmsflexF=0.5/{\rm arcmin}, \rmsflexG=0.9/{\rm arcmin}$}

Assuming that our noise is dominated by Poisson noise from intrinsic galaxy shapes, the noise power spectra are
\bea
P_\gamma^{\rm noise}(l) &=& \rmsshear^2/\ngal \\
P_\flexF^{\rm noise}(l) &=& \rmsflexF^2/\ngal \\
P_\flexG^{\rm noise}(l) &=& \rmsflexG^2/\ngal 
\eea
where $\ngal$ is the projected number density of galaxies and $\rmsshear$, $\rmsflexF$, or $\rmsflexG$ is the RMS shape for a single galaxy.  If we reconstruct the convergence from a minimum-variance-weighted linear combination of the shear and flexion \citep[as in][]{okura_umetsu_etal_2007}, then the noise power spectrum of our convergence map will be
\beq \label{eq:kappanoise}
P_\kappa^{\rm noise}(l) = \left( \frac{1}{P_\gamma^{\rm noise}}
 + \frac{l^2}{P_\flexF^{\rm noise}} + \frac{l^2}{P_\flexG^{\rm noise}} \right)^{-1}
.\eeq
The different $l$-dependencies of shear and flexion noise are explained in Appendix \ref{sec:flex_apx}.  For shape maps smoothed by a gaussian filter, the shape noise contribution to $\mean{\kappamap^2}$ is
\beq \label{eq:cp}
C_p(\smoothangle) = \frac{1}{2\pi}\int_0^\infty\dx{l}l \tilde F_ \smoothangle(l)^2P_\kappa^{\rm noise}(l)
\eeq
where the lower limit of integration, $l=0$, implies a very large map (we will consider small maps presently).  Substituting \refeq{kappanoise} into \refeq{cp} and defining
\beq
\sigflex^{-2} \equiv \rmsflexF^{-2} + \rmsflexG^{-2}
\eeq
we find
\bea
\label{eq:pixnoiseplugin}
C_p(\smoothangle) &=& \frac{1}{2\pi}\int_0^\infty\frac{dl}{l}\left(\frac{1}{l^2\rmsshear^2}+\frac{1}{\sigflex^2}\right)^{-1} (e^{-\frac{1}{8}l^2\smoothangle^2})^2 \\
	&=& \frac{\sigflex^2}{4\pi\ngal}\exp{\left(\frac{\smoothangle^2\sigflex^2}{4\rmsshear^2}\right)} \Gamma{\left(0,\frac{\smoothangle^2\sigflex^2}{4\rmsshear^2}\right)}
\eea
where $\Gamma(s,x)$ is the incomplete gamma function
\beq
\Gamma(s, x) \equiv \int_x^\infty \dx{t} e^{-t}t^{s-1}
\eeqp

In the shear-only limit, $\sigflex \rightarrow \infty$, the property
\beq
\lim_{x\rightarrow\infty} \Gamma(s, x)e^x x^{1-s}=1
\eeq
leads to $C_p(\smoothangle)\approx\rmsshear^2/(\pi\smoothangle^2\ngal)$, as expected for a KS inversion \citep{seitz_schneider_1995}.  Decreasing the smoothing scale, $\smoothangle\rightarrow 0$, we see that small-scale noise leads to a divergence in the shear-only limit; hence we must smooth to obtain finite noise.  In the flexion-only limit, $\rmsshear\rightarrow\infty$, $C_p(\smoothangle)$ diverges due to large-scale noise, which can be seen by ignoring the shear term in \refeq{pixnoiseplugin}.  For finite maps, there are no noise contributions from modes larger than the map, so
\beq
\mean{\nu_\smoothangle(\skypos;\fieldsize)^2} = C_p(\smoothangle) - C_p(\fieldsize)
\eeqp
Although we are considering a tophat-shaped map, and the excluded noise $C_p(\fieldsize)$ assumes a large {\em Gaussian} filter, we note that for large enough $\fieldsize$, we recover the shear-only limit, which is the same for either filter function.

When shear and flexion are combined, they attenuate shape noise in a complementary way.  We can see in \refeq{pixnoiseplugin} that the shear term prevents a divergence toward small $l$ while the flexion term {\em reduces} the divergence toward large $l$.  Without the smoothing factor, there is an ultraviolet divergence even with flexion, but it is less severe than for shear.  The benefit of a combined approach is illustrated in \reffig{shapenoise}, which compares shape noise from a shear-only inversion to one that includes flexion.  Since flexion suppresses small-scale noise, we can smooth our shape maps on smaller angular scales without significantly increasing the noise in our convergence map. Thus, our convergence map will recover more small fluctuations, leading to a larger $\mean{\kappasignal\kappaBHB}$ and therefore a larger $r^2$ by \refeq{rdef_terms}.  Of course, this improvement degrades as we increase $\rmsflexF$ and $\rmsflexG$.

\section{Results} \label{results}

For all calculations, we adopt a flat, \LCDM model with $h=0.7$, $\Omega_m=0.27$, $\Omega_b=0.045$, $\sigma_8=0.8$ and $n_s=0.96$, consistent with the WMAP 5-year parameters \citep{dunkley_komatsu_etal_2008}.  We calculate the linear matter power spectrum using the fitting formula of Eisenstein \& Hu (\citeyear{eisenstein_hu_1999}) without baryon wiggles, and we apply to this the nonlinear fitting formula of Smith et al.\  (\citeyear{smith_peacock_etal_2003}).  The weak lensing fields are assumed to be Gaussian and uncorrelated on different scales.  For our cosmological model, we expect that without delensing, the uncertainty in the lensing of a SMBHB, $\sigma(\kappa)=\sigma(\mu)/2$, will be 3.9\% for $z_\BHB=2$ and 5.2\% for $z_\BHB=3$.  By \refeq{magdist} this uncertainty is equal to the relative distance error, $\sigma(D_L)/D_L$.

\subsection{Distance error reduction from a deep image}

We first consider delensing with a narrow, deep image of a similar size and depth to the Hubble Ultra Deep Field (HUDF).  The HUDF contains over 8000 galaxies detected at over $10\sigma$ within an area of 12 arcmin$^2$ \citep{coe_benitez_etal_2006}.  We take our galaxy redshift distribution to be
\beq \label{eq:Smail}
\pgal(z) \propto z^\alpha e^{-(z/z0)^\beta}
\eeq
with $\alpha=0.8,\beta=2.0$ and $z_{\rm med}=1.8$ in order to approximate the redshift distribution of the HUDF.  However, we suppose that we have the resolution provided by adaptive optics with an Extremely Large Telescope (ELT) of 50mas \citep{elt_science_2006}.  We estimate that such an instrument will provide RMS galaxy shapes of 
\beq
\rmsshear=0.2 \eqand \rmsflexF=\frac{0.5}{{\rm arcmin}} \eqand \rmsflexG=\frac{0.9}{{\rm arcmin}}
\eeq
where the covariance between these measures is negligible \citep[][and Rowe et al. 2009 in preparation]{massey_rowe_etal_2007}.  With an ELT, we can consider going deeper than the HUDF so as to obtain a denser galaxy field.  However, since galaxies are roughly 1 arcsec$^{2}$ in size, we will become confusion-limited as we approach $\ngal\sim 1/{\rm arcsec}^{2}=3600/{\rm arcmin}^{2}$.

As D03 point out, there is generally an optimal smoothing angle for a shear map.  If the smoothing is too fine, the map will be very noisy; if smoothing is too coarse, the small-scale magnification fluctuations will be washed away.  When adequate flexion measurements are included with the shears, noise can be reduced enough to make the optimal smoothing scale smaller than the inter-galaxy spacing.  Of course, it makes no sense to smooth on scales smaller than the inter-galaxy spacing; we therefore choose a conservative cutoff of
\beq
\ngal \pi \smoothangle^2 = 10
\eeqp
For example, for a galaxy density of $\ngal=800/{\rm arcmin}^{2}$, we would use $\smoothangle=0.063$ arcmin.  \reffig{shapenoise} shows that at this scale, including flexion reduces the variance in the noise by more than an order of magnitude relative to shear alone.

We define the distance error remainder, $\remain$, to be the SMBHB distance error after delensing divided by the original distance error:
\beq
\remain \equiv \frac{\sigma^\prime(D_L)}{\sigma(D_L)}
\eeqp
Since $\sigma(D_L)\propto\sigma(\kappa)$, the equation for $\remain$ follows directly from \refeq{rdef}:
\beq
\remain = \sqrt{1-r^2}
\eeqp
\reffig{ELT_deep_contour_z2} contains contour plots showing the $\remain$ we could obtain from shape maps of a given size and galaxy density.  The solid contours assume a HUDF-like galaxy redshift distribution, given by \refeq{Smail}, while the dashed contours assume that all galaxies lie in a plane at the redshift of the SMBHB.  The latter configuration is clearly a best-case scenario and provides an upper limit on how well a map can measure the SMBHB magnification.

As an example, consider the plot with $z_\BHB=2$ in \reffig{ELT_deep_contour_z2}.  Going only by the solid contours, the ``HUDF'' label lies near $\remain=0.7$.  Therefore we could use an HUDF-like ELT image to reduce the SMBHB distance uncertainty to about 70\% of its original value (i.e. we reduce the error bar by 30\%); this assumes that we use no redshift information about individual galaxies and do not attempt to break the mass-sheet degeneracy.  In this case, the remaining uncertainty comes primarily from convergence fluctuations on scales larger than our map.  While we can improve our measurement of the SMBHB magnification by widening our map, there are diminishing returns on making it deeper.  With substantial access to an ELT we could create a mosaic to reduce our distance error to less than 60\% of the original value.

We reiterate that the solid contours in \reffig{ELT_deep_contour_z2} assume that we only construct 2D weak lensing maps.  If we can measure the redshifts of individual galaxies and down-weight galaxies as they deviate from $z_\BHB$, we could improve on these results.  The $z_\BHB=3$ plot highlights the importance of the galaxy redshift distribution.  Because the galaxies have $\zmed=1.8$, most of them are closer to us than the SMBHB and are therefore lensed by different structures.  Hence, there is a large discrepancy between the solid (realistic) and dashed (idealized) contours.  Meanwhile, the galaxies are almost evenly distributed around a SMBHB at $z_\BHB=2$, so there is a smaller difference between the contour sets.

\Sfigtwo{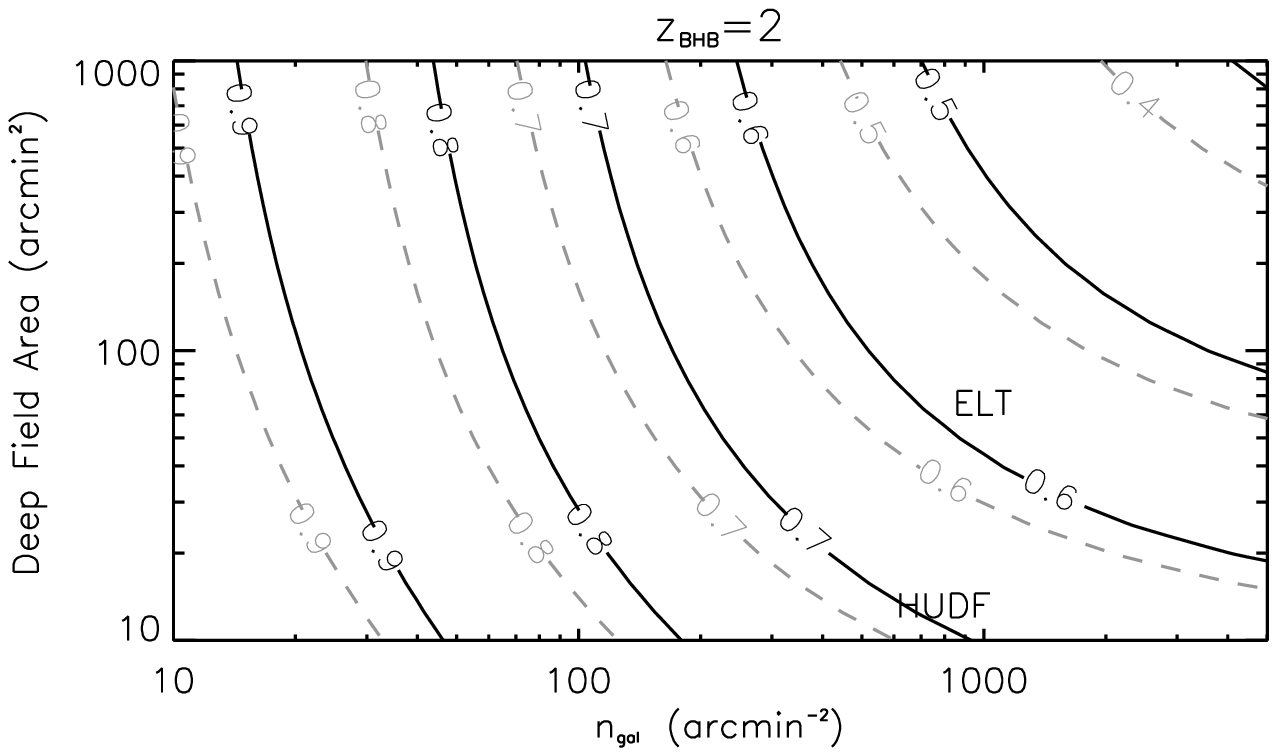}{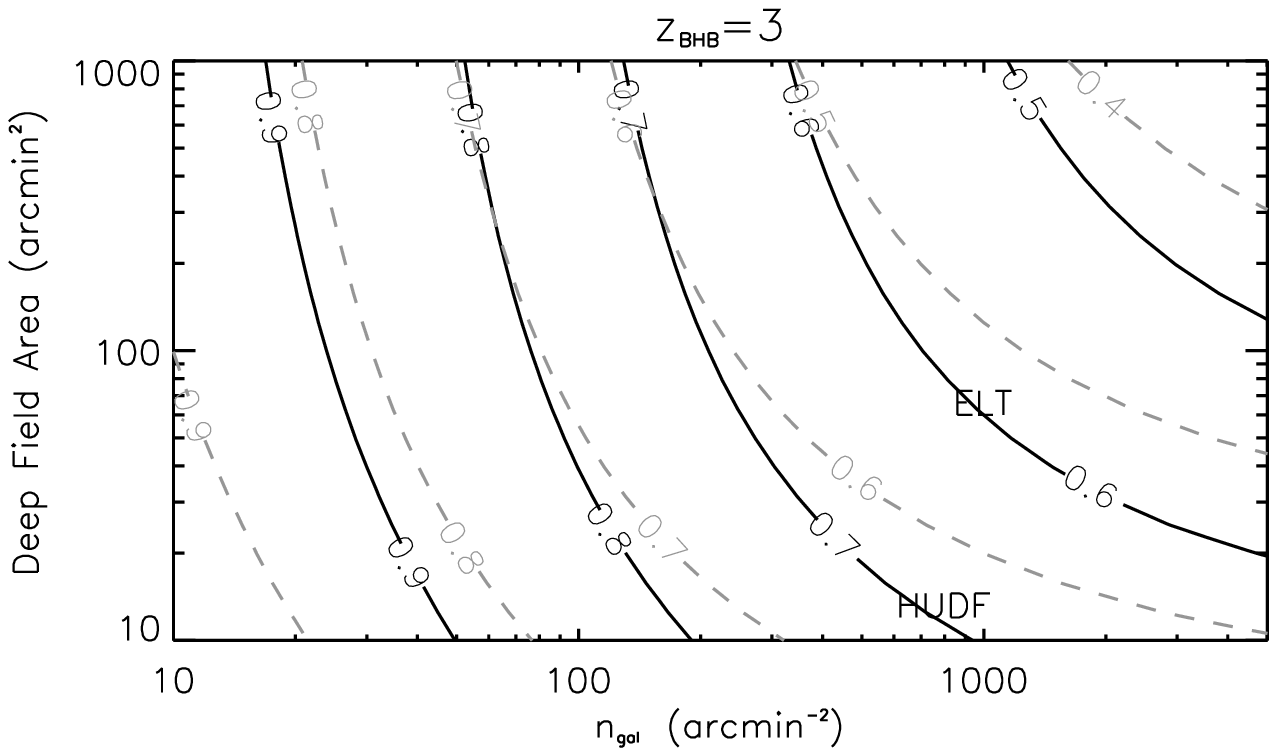}{Contours of $\remain$, the fraction of distance error remaining after measuring a SMBHB's magnification with a narrow, deep image.  Solid lines assume the galaxy redshift distribution given by \refeq{Smail}.  Dashed lines assume that all galaxies have the same redshift as the SMBHB.  The point marked ``HUDF'' is for an ELT image similar in size and depth to the Hubble Ultra Deep Field.  The point marked ``ELT'' assumes a large tiled image from an ELT.}

\subsection{Distance error reduction from hybrid maps}
\label{results_hybrid}

It is apparent from \reffig{ELT_deep_contour_z2} that we could significantly improve our convergence reconstruction with wider images around each SMBHB.  Unfortunately, it would be impractical to use telescopes such as an ELT to create mosaics larger than a few tens of square arcminutes.  A space survey telescope such as JDEM \citep{jdem_web} or Euclid \citep{euclid_science_2008} would provide the required width but it would be blind to sub-arcminute convergence fluctuations that require high galaxy densities.  We therefore propose making convergence maps from a hybrid of deep, pointed observations and wider survey images.  The idea here is that the deep images will be used to measure small-angle convergence modes, while the wide images will pick up modes larger than the size of the deep images.

Accordingly, we relabel $\kappamap$ to $\kappadeep$, and we let $\kappawide$ denote the convergence map obtained from the larger survey.  We are unconcerned with small fluctuations outside of the deep map.  Therefore to minimize shape noise in the wide map, we smooth it by a top-hat filter of radius $\fieldsize$, the size of the deep map:
\beq
\kappadeep(\skypos) = \kappasignal(\skypos) - \kappaMS(\skypos) + \noise_\smoothangle(\skypos;\fieldsize)
\eeq
\beq
\kappawide(\skypos) = \kappaMS^\star(\skypos) + \noise^\star_\fieldsize(\skypos)
\eeqp
Because each of these terms depends on the galaxy density and redshift distributions of the corresponding images, we have introduced the $^\star$ superscripts to signify a dependance on the $\ngal$ and $\pgal(z)$ of the wide survey.  For a sufficiently wide survey, the shape noise at $\skypos$ is independent of the survey area, so we make that approximation here.

We can split $\kappaBHB$ into terms comprised of modes with wavelengths that are smaller or larger than the narrow, deep image area:
\beq
\kappaBHB = \kappa_{\rm S}+ \kappa_{\rm L}
\eeq
The subscripts denote the ``small'' and ``large'' terms.  As in \refeq{MSconst}, we estimate the large term to be equal to the total convergence smoothed on the scale of the deep map:
\beq
\kappa_{\rm L} \approx {\kappaBHB}_{;\fieldsize}
\eeq
We now define ``small'' and ``large'' versions of $r^2$:
\beq
r_{\rm S}^2 \equiv \frac{\mean{\kappadeep\kappa_{\rm S}}^2}{\mean{\kappa_{\rm S}^2}\mean{\kappadeep^2}}
	= \frac{\mean{(\kappasignal-\kappaMS)\kappa_{\rm S}}^2}
	{\mean{\kappa_{\rm S}^2}\mean{\kappasignal^2-\kappaMS^2+\noise_\smoothangle(\fieldsize)^2}}
\eeq
\beq
r_{\rm L}^2 \equiv \frac{\mean{\kappawide\kappa_{\rm L}}^2}{\mean{\kappa_{\rm L}^2}\mean{\kappawide^2}}
	= \frac{\mean{\kappaMS^\star\kappa_{\rm L}}^2}
	{\mean{\kappa_{\rm L}^2}\mean{{\kappaMS^\star}^2+{\noise^\star_\fieldsize}^2}}
\eeqp
Our final uncertainty in $\kappaBHB$ is now given by 
\beq
\sigma^\prime(\kappaBHB)^2 = 
 (1-r_{\rm deep}^2) \mean{\kappa_{\rm S}^2} + (1-r_{\rm wide}^2) \mean{\kappa_{\rm L}^2} 
\eeqp

\Sfigtwo{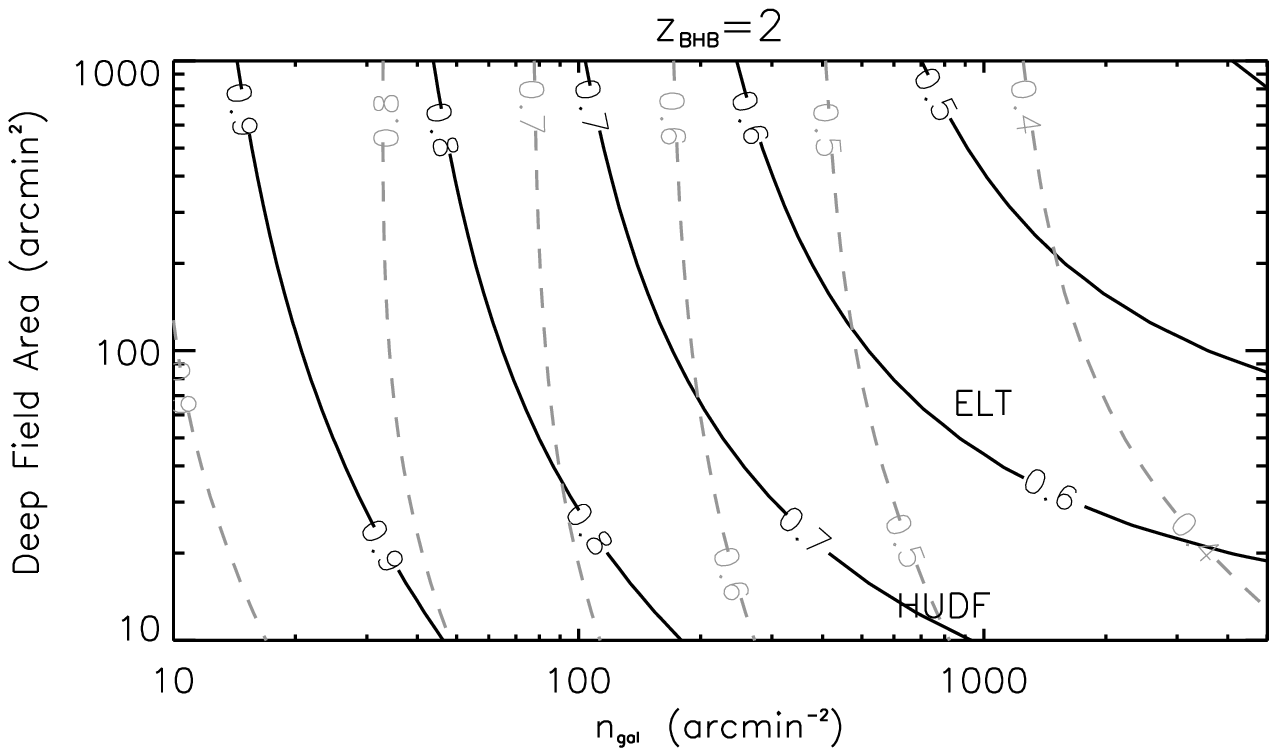}{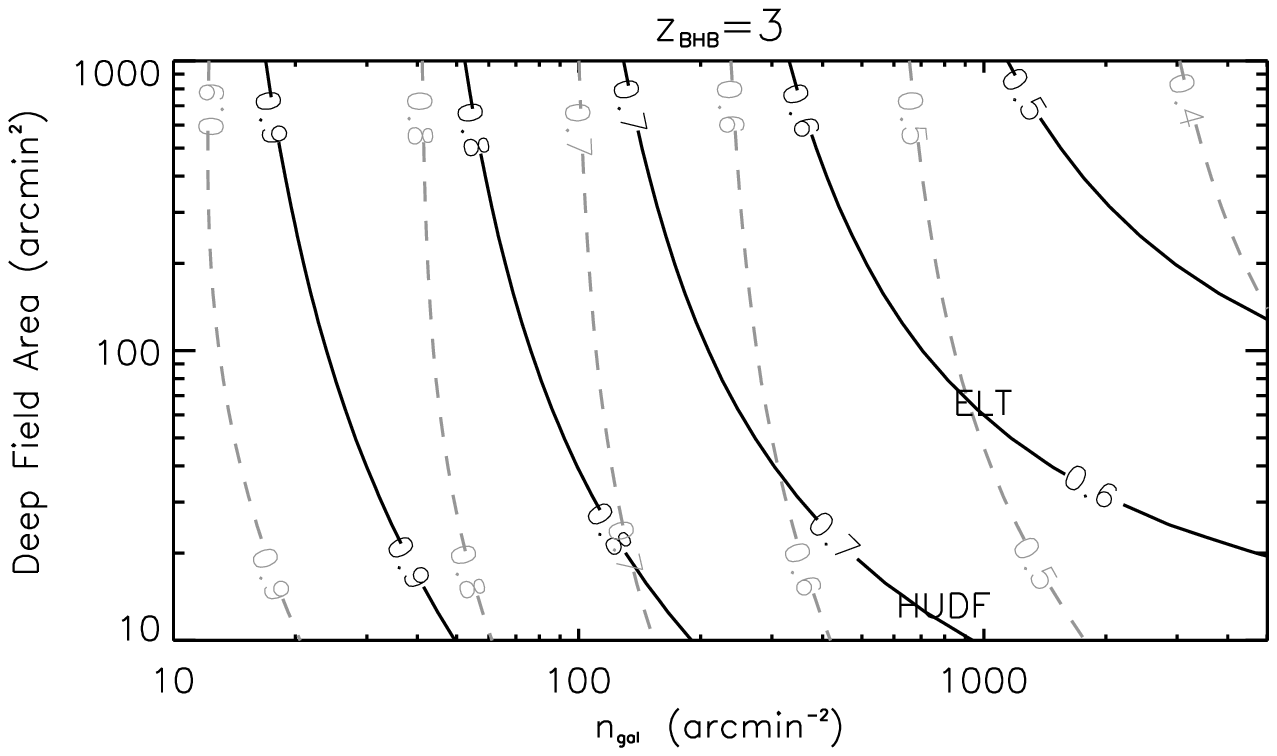}{Solid contours are the same as in \reffig{ELT_deep_contour_z2}.  Dashed contours show the improvement when a shear map from a survey space telescope is used in tandem with an ELT.}

We take our space-based survey to have $\ngal=100$ and a redshift distribution given by  \refeq{Smail} with $\alpha=2$, $\beta=1.5$ and $\zmed=1.5$ \citep{albrecht_bernstein_etal_2006}.  We take the intrinsic shear to be $\rmsshear=0.25$, but we do not consider flexion measurements from the wide survey.  Flexion measurements will have larger errors ($\simeq 1$ per arcmin)  with a survey instrument, and since flexion's sensitivity to matter fluctuations decreases with angular size, it should only marginally improve a shear measurement of $\kappa_{\rm L}$.  \reffig{ELT_hybrid_contour_z2} shows how including a wide map improves on a deep map's ability to delens a SMBHB.  For $z_\BHB=2$, delensing with a HUDF-like image by itself can reduce the distance error by about 30\%, but when combined with the space survey, half of the uncertainty can be removed.  The improvement is slightly less for $z_\BHB=3$; we should expect this since most galaxies in the space survey have a redshift lower than 3.  Notice now that creating a large ELT mosaic provides a little improvement but will not be as effective as making the image deeper.  Again, these results assume that we use 2D maps and do not attempt to break the mass-sheet degeneracy with information beyond shape measurements.

\Sfig{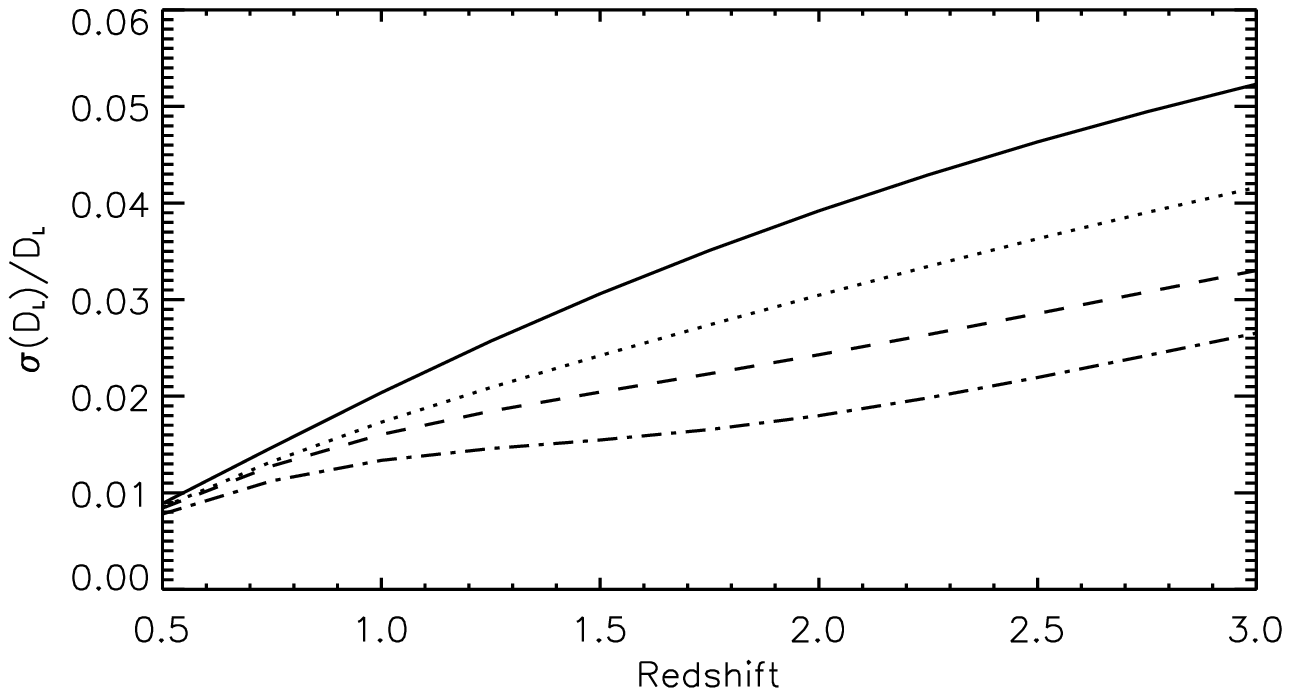}{Uncertainty in SMBHB distance versus SMBHB redshift, with and without delensing.  Solid: distance uncertainty without delensing.  Dotted: reduction expected after shear-only delensing with a 30 arcmin$^2$ image from an ELT ($\ngal=1000/{\rm arcmin}^2$, $\zmed=1.8$).  Dashed: shear and flexion delensing with an ELT.  Dot-dashed: shear and flexion delensing with an ELT combined with a space survey telescope.  2D shear and flexion maps are assumed (no individual redshift measurements).
Note that this ELT image is smaller than the one marked in Figures \ref{fig:ELT_deep_contour_z2} and \ref{fig:ELT_hybrid_contour_z2}.}

\section{Conclusions and Future Work} \label{conclusions}

Gravitational lensing of supermassive black hole binary systems (SMBHBs) severely limits their usefulness as standard sirens -- precise distance probes based on known gravitational wave signals.  We suggest that ``delensing'' each SMBHB -- i.e. mapping the magnification field around it using a Kaiser-Squires-like inversion technique -- could substantially reduce distance errors due to lensing.  \reffig{error_improve_ELT} illustrates the error reduction that could be achieved with a 30 arcmin$^2$ image from an Extremely Large Telescope combined with a space-based survey such as JDEM or Euclid.  We find that combining 2D maps from these instruments could reduce SMBHB distance error from lensing by about a factor of 2 for $z_\BHB > 1.5$.  The success of SMBHB delensing requires adequate flexion measurements to recover fine fluctuations in the magnification field and shear measurements to recover larger features.

It may be possible to achieve superior results by using techniques beyond the simple Kaiser-Squires inversion considered here.  For instance, incorporating a redshift for each galaxy would enable some improvement over 2D maps.  Redshifts could also help break the mass-sheet degeneracy in a deep image, thereby eliminating some of the need for a wide survey.  The degeneracy could also be broken by including magnification maps from quasar number counts.  Furthermore, with the advent of an ELT, improved high-resolution shape measurement techniques may emerge, leading to maps with significantly lower shape noise.

We have made several simplifying assumptions that should be addressed in future work.  We have extended the Smith et al.\  fitting formula for the matter power spectrum beyond its range of accuracy, and we have assumed that matter fluctuations remain Gaussian random.  A more complete treatment of the matter power spectrum, e.g. from the results of N-body simulations, will be needed to refine the technique of delensing and estimate its potential success (cf. Gair, King, et al.\ in preparation).  It may be possible to exploit the fact that non-linear gravitational clustering causes the true weak lensing fields to contain less information than their Gaussian random approximations \citep{cooray_hu_2001,rimes_hamilton_2005, dore_lu_etal_2009}.  Finally, we have ignored higher order weak lensing effects such as reduced shear/flexion and contaminating effects such as intrinsic alignments.





\section*{Acknowledgments}

Thanks to Daniel Holz, Hiranya Peiris, and Daniel Thomas for valuable discussions.  Calculations were done in part by modifying the iCosmo IDL package \citep{icosmo_2008}.  CS is funded by an STFC Rolling Grant.  DB acknowledges an STFC Advanced Fellowship and an RCUK Research Fellowship.  

\appendix

\section{Effect of the mass-sheet degeneracy on delensing}
\label{mass-sheet}

Let $\kappasignal$ be the true convergence of our galaxy sample smoothed by a filter with a characteristic angular scale $\smoothangle$.  Ignoring shape noise, our convergence map, $\kappamap$, is related to the true convergence by 
\beq \label{eq:MSdef}
\kappamap = (1-\beta)\kappasignal +\beta
\eeq
for some unknown constant $\beta$.  This is the mass-sheet degeneracy: shape measurements cannot distinguish the scaling factor $(1-\beta)$ from a constant sheet of mass with convergence $\beta$.  Let us think of $\kappasignal$ as being composed of two terms:
\beq
\kappasignal = {\kappasignal}_S + {\kappasignal}_L 
\eeq
where ${\kappasignal}_S$ (or ${\kappasignal}_L$) is a sum over Fourier modes with wavelengths smaller (larger) than the scale of the map.  Note that if we restrict our attention to the area of the map, we can treat ${\kappasignal}_L$ as a constant, which we call $K(\fieldsize)$ for a circular map of radius $\fieldsize$.

If we apply an initial transformation \refeq{MSdef} to $\kappamap$ to obtain a new map with $\mean{\kappamap}=0$, then this new map should be proportional to ${\kappasignal}_S$ and still related to $\kappasignal$ by \refeq{MSdef}.  Therefore, since $\kappamap=0$ when ${\kappasignal}_S=0$, we have
\beq
1-\beta = -\beta/K(\fieldsize) = [1-K(\fieldsize)]^{-1}
\eeqp
We can now write
\beq
\kappamap = \frac{{\kappasignal}_S}{1-K(\fieldsize)} = \frac{\kappasignal-K(\fieldsize)}{1-K(\fieldsize)}
\eeq
Thus, our map basically allows us to reconstruct ${\kappasignal}_S$ up to some multiplicative constant which is determined by fluctuations larger than the map.  For weak distortions ($\kappa \ll 1$), we can ignore this factor to first order:
\beq
\kappamap \approx {\kappasignal}_S[1+K(\fieldsize)] \approx {\kappasignal}_S=\kappasignal-K(\fieldsize)
\eeqp
So to first order, our map determines the total convergence field modulo some unknown additive constant, $K(\fieldsize)$.  Delensing reduces our uncertainty in ${\kappasignal}_S$ but not $K(\fieldsize)$, which cannot be determined from shape measurements within the map.  Measuring $K(\fieldsize)$ requires that we either widen the map or else break the mass-sheet degeneracy using additional information in the field (e.g. redshifts).

The higher order corrections can be significant when calculating 2-point statistics.  For instance, at second order, the term $\mean{\kappamap\kappaBHB}$ in \refeq{rdef_terms} becomes
\beq
\mean{\kappamap\kappaBHB} \rightarrow \mean{\kappamap\kappaBHB} + \mean{\kappamap\kappaBHB\kappaMS}
\eeq
The 3rd order term vanishes for purely Gaussian fields, but in reality such corrections can be as much as 10\%, depending on the amplitude of the non-linear matter power spectrum on small scales \citep{white_2005,shapiro_2009}.  We expect the particular correction above to be partially suppressed due to the low correlation between $\kappamap$ and $\kappaMS$, which represent different angular scales.  Such non-Gaussian/non-linear effects will need to be accounted for in a realistic delensing scheme, but they are beyond the accuracy that we are interested in for this paper. 

\section{Shape noise in a Kaiser-Squires inversion with shear and flexion} \label{sec:flex_apx}

A Kaiser-Squires-like inversion of shape measurements to reconstruct the convergence field involves the following steps:
\begin{enumerate}
\item Measure shear and flexion ($\gamma$, $\flexF$, $\flexG$) for each galaxy
\item Smooth each shape map with a filter function $F_\smoothangle(\skypos)$
\item Fourier transform each smoothed map
\item Convert shear and flexion to convergence in Fourier space\label{convert}
\item Calculate convergence modes using minimum variance weighting
\item Inverse Fourier transform to obtain $\kappamap(\skypos)$.
\end{enumerate}
Because shear and flexion are converted to convergence differently in step \ref{convert}, shape noise in the resulting $\kappa$ map depends differently on intrinsic shears and flexions.  We now briefly review the conversion equations.

We start by introducing the complex notation for shear and flexion:
\bea
\FTshear(\vl) &=& \FTshear_1(\vl) + i \FTshear_2(\vl) \\
\FTflexF(\vl) &=& \FTflexF_1(\vl) + i \FTflexF_2(\vl) \\
\FTflexG(\vl) &=& \FTflexG_1(\vl) + i \FTflexG_2(\vl)
\eea
Next, borrowing notation from \citet{bartelmann_schneider_2001}, we define the following complex kernels in Fourier space:
\bea
\kkern_\gamma(\vl) &\equiv& \frac{\pi}{l^2}(l_1+il_2)^2 \\
\kkern_\flexF(\vl) &\equiv& -\frac{i\pi}{l^2}(l_1+il_2)  \\
\kkern_\flexG(\vl) &\equiv& \frac{i\pi}{l^4}(l_1+il_2)^3
\eea
where $l\equiv|\vl|\equiv l_1^2+l_2^2$. Note that the subscripts on $l$ denote the $x$- and $y$-components of the wave-vector $\vl$ relative to some fixed $x$-axis, while subscripts on shear or flexion denote polarizations defined by the same $x$-axis.  Hereafter, we will not explicitly write the dependence of quantities on $\vl$.

In Fourier space, for $\vl\neq0$, weak lensing distortions are related to the convergence by
\footnote{N.B. equations (62) and (63) in \citet{bacon_goldberg_etal_2006} are missing a factor of 1/2 on the right-hand side, cf. their equation (13).}
\citep{bacon_goldberg_etal_2006}
\bea
\FTshear &=& \pi^{-1}\FTkappa\,\kkern_\gamma \\
\FTflexF &=& \pi^{-1}\FTkappa\,\kkern_\flexF l^2 \\
\FTflexG &=& \pi^{-1}\FTkappa\,\kkern_\flexG l^2
\eea
We can therefore use the fact that
\beq
\kkern_\gamma^* \kkern_\gamma = \pi^2
\eeq
and
\beq
\kkern_\flexF^* \kkern_\flexF = \kkern_\flexG^* \kkern_\flexG = \pi^2 l^{-2}
\eeq
to convert shape modes to convergence modes using the following equations:
\bea
\label{eq:s2k} \tilde\kappamap &=& \pi^{-1}\kkern_\gamma^*\FTshear \\
\label{eq:F2k} \tilde\kappamap &=& \pi^{-1}\kkern_\flexF^*\FTflexF \\
\label{eq:G2k} \tilde\kappamap &=& \pi^{-1}\kkern_\flexG^*\FTflexG
\eea

The convergence power spectrum of our $\kappa$ map is defined by
\beq
\mean{\tilde\kappamap(\vl)\tilde\kappamap(\vl^\prime)^*} = (2\pi)^2 \delta(\vl-\vl^\prime) P_{\kappamap}(l)
\eeqc
with similar definitions for the shear and flexion maps.  Using \refeq{s2k}, \refeq{F2k} and \refeq{G2k}, it is easy to show that a convergence power spectrum obtained from inverting the white noise spectrum of intrinsic shear or flexion will contain a corresponding factor of $\kkern\kkern^*$.  The result is
\bea
P_\kappa^{\rm noise}(l) &=& P_\gamma^{\rm noise} \\
P_\kappa^{\rm noise}(l) &=& P_\flexF^{\rm noise} l^{-2} \\
P_\kappa^{\rm noise}(l) &=& P_\flexG^{\rm noise} l^{-2}
. \eea
If we compute each $\tilde\kappa$ from a linear combination of shear and flexion using minimum variance weighting, the final noise spectrum is given by
\beq
\frac{1}{P_\kappa^{\rm noise}(l)} = \frac{1}{P_\gamma^{\rm noise}} + \frac{l^2}{P_\flexF^{\rm noise}} + \frac{l^2}{P_\flexG^{\rm noise}}
\eeqp

\label{lastpage}

\bibliography{flexion_delensing,master_library}

\begin{thebibliography}{46}
\expandafter\ifx\csname natexlab\endcsname\relax\def\natexlab#1{#1}\fi

\bibitem[{{Albrecht} {et~al.}(2006){Albrecht}, {Bernstein}, {Cahn}, {Freedman},
  {Hewitt}, {Hu}, {Huth}, {Kamionkowski}, {Kolb}, {Knox}, {Mather}, {Staggs},
  \& {Suntzeff}}]{albrecht_bernstein_etal_2006}
{Albrecht}, A. {et~al.} 2006, ArXiv e-prints, astro-ph/0609591

\bibitem[{{Arun} {et~al.}(2009){Arun}, {Babak}, {Berti}, {Cornish}, {Cutler},
  {Gair}, {Hughes}, {Iyer}, {Lang}, {Mandel}, {Porter}, {Sathyaprakash},
  {Sinha}, {Sintes}, {Trias}, {Van Den Broeck}, \&
  {Volonteri}}]{arun_babak_etal_2009}
{Arun}, K.~G. {et~al.} 2009, Classical and Quantum Gravity, 26, 094027,
  0811.1011

\bibitem[{{Bacon} {et~al.}(2006){Bacon}, {Goldberg}, {Rowe}, \&
  {Taylor}}]{bacon_goldberg_etal_2006}
{Bacon}, D.~J., {Goldberg}, D.~M., {Rowe}, B.~T.~P., \& {Taylor}, A.~N. 2006,
  \mnras, 365, 414, arXiv:astro-ph/0504478

\bibitem[{{Bartelmann} \& {Schneider}(2001)}]{bartelmann_schneider_2001}
{Bartelmann}, M., \& {Schneider}, P. 2001, \physrep, 340, 291,
  arXiv:astro-ph/9912508

\bibitem[{{Bender} {et~al.}(1994)}]{bender_etal_1994}
{Bender}, P., {et~al.} 1994, LISA, Laser interferometer space antenna for
  gravitational wave measurements: ESA Assessment Study Report

\bibitem[{{Bower}(1991)}]{bower_1991}
{Bower}, R.~G. 1991, \mnras, 248, 332

\bibitem[{Coe {et~al.}(2006)Coe, Benitez, Sanchez, Jee, Bouwens, \&
  Ford}]{coe_benitez_etal_2006}
Coe, D., Benitez, N., Sanchez, S.~F., Jee, M., Bouwens, R., \& Ford, H. 2006,
  The Astronomical Journal, 132, 926

\bibitem[{{Cooray} \& {Hu}(2001)}]{cooray_hu_2001}
{Cooray}, A., \& {Hu}, W. 2001, \apj, 554, 56, arXiv:astro-ph/0012087

\bibitem[{{Crittenden} {et~al.}(2001){Crittenden}, {Natarajan}, {Pen}, \&
  {Theuns}}]{crittenden_natarajan_etal_2001}
{Crittenden}, R.~G., {Natarajan}, P., {Pen}, U.-L., \& {Theuns}, T. 2001, \apj,
  559, 552, arXiv:astro-ph/0009052

\bibitem[{{Dalal} {et~al.}(2003){Dalal}, {Holz}, {Chen}, \&
  {Frieman}}]{dalal_holz_etal_2003}
{Dalal}, N., {Holz}, D.~E., {Chen}, X., \& {Frieman}, J.~A. 2003, \apjl, 585,
  L11, arXiv:astro-ph/0206339

\bibitem[{{Dalal} {et~al.}(2006){Dalal}, {Holz}, {Hughes}, \&
  {Jain}}]{dalal_holz_etal_2006}
{Dalal}, N., {Holz}, D.~E., {Hughes}, S.~A., \& {Jain}, B. 2006, \prd, 74,
  063006, arXiv:astro-ph/0601275

\bibitem[{Dodelson {et~al.}(2006)Dodelson, Shapiro, \&
  White}]{dodelson_shapiro_etal_2006}
Dodelson, S., Shapiro, C., \& White, Martin~J., . 2006, Phys. Rev., D73,
  023009, astro-ph/0508296

\bibitem[{{Dor{\'e}} {et~al.}(2009){Dor{\'e}}, {Lu}, \&
  {Pen}}]{dore_lu_etal_2009}
{Dor{\'e}}, O., {Lu}, T., \& {Pen}, U.-L. 2009, ArXiv e-prints, 0905.0501

\bibitem[{{Dunkley} {et~al.}(2008){Dunkley}, {Komatsu}, {Nolta}, {Spergel},
  {Larson}, {Hinshaw}, {Page}, {Bennett}, {Gold}, {Jarosik}, {Weiland},
  {Halpern}, {Hill}, {Kogut}, {Limon}, {Meyer}, {Tucker}, {Wollack}, \&
  {Wright}}]{dunkley_komatsu_etal_2008}
{Dunkley}, J. {et~al.} 2008, ArXiv e-prints, 803, 0803.0586

\bibitem[{{Eichler} {et~al.}(1989){Eichler}, {Livio}, {Piran}, \&
  {Schramm}}]{eichler_livio_etal_1989}
{Eichler}, D., {Livio}, M., {Piran}, T., \& {Schramm}, D.~N. 1989, \nat, 340,
  126

\bibitem[{{Eisenstein} \& {Hu}(1999)}]{eisenstein_hu_1999}
{Eisenstein}, D.~J., \& {Hu}, W. 1999, \apj, 511, 5, arXiv:astro-ph/9710252

\bibitem[{{ELT Science Working Group}(2006)}]{elt_science_2006}
{ELT Science Working Group}. 2006, Report of the ELT Science Working Group,
  Available 6 July 2009 @ http://www.eso.org/sci/facilities/eelt/science/doc/

\bibitem[{{Heymans} \& {Heavens}(2003)}]{heymans_heavens_2003}
{Heymans}, C., \& {Heavens}, A. 2003, ArXiv e-prints, astro-ph/0310495

\bibitem[{{Hirata} \& {Seljak}(2004)}]{hirata_seljak_2004}
{Hirata}, C.~M., \& {Seljak}, U. 2004, \prd, 70, 063526, arXiv:astro-ph/0406275

\bibitem[{Holz \& Hughes(2005)}]{holz_hughes_2005}
Holz, D.~E., \& Hughes, S.~A. 2005, \apj, 629, 15

\bibitem[{{J{\"o}nsson} {et~al.}(2005){J{\"o}nsson}, {Dahlen}, {Goobar},
  {Gunnarsson}, {M{\"o}rtsell}, \& {Lee}}]{jonsson_dahlen_etal_2005}
{J{\"o}nsson}, J., {Dahlen}, T., {Goobar}, A., {Gunnarsson}, C.,
  {M{\"o}rtsell}, E., \& {Lee}, K. 2005, in Bulletin of the American
  Astronomical Society, Vol.~37, 1459--+

\bibitem[{{J{\"o}nsson} {et~al.}(2006{\natexlab{a}}){J{\"o}nsson},
  {Dahl{\'e}n}, {Goobar}, {Gunnarsson}, {M{\"o}rtsell}, \&
  {Lee}}]{jonsson_dahlen_etal_2006a}
{J{\"o}nsson}, J., {Dahl{\'e}n}, T., {Goobar}, A., {Gunnarsson}, C.,
  {M{\"o}rtsell}, E., \& {Lee}, K. 2006{\natexlab{a}}, \apj, 639, 991,
  arXiv:astro-ph/0506765

\bibitem[{{J{\"o}nsson} {et~al.}(2006{\natexlab{b}}){J{\"o}nsson},
  {Dahl{\'e}n}, {Goobar}, {Gunnarsson}, {M{\"o}rtsell}, \&
  {Lee}}]{jonsson_dahlen_etal_2006}
------. 2006{\natexlab{b}}, \apj, 639, 991, arXiv:astro-ph/0506765

\bibitem[{{J{\"o}nsson} {et~al.}(2007){J{\"o}nsson}, {Goobar}, \&
  {M{\"o}rtsell}}]{jonsson_goobar_etal_2007}
{J{\"o}nsson}, J., {Goobar}, A., \& {M{\"o}rtsell}, E. 2007, \apj, 658, 52,
  arXiv:astro-ph/0611334

\bibitem[{{J{\"o}nsson} {et~al.}(2009){J{\"o}nsson}, {M{\"o}rtsell}, \&
  {Sollerman}}]{jonsson_mortsell_etal_2009}
{J{\"o}nsson}, J., {M{\"o}rtsell}, E., \& {Sollerman}, J. 2009, \aap, 493, 331

\bibitem[{Kaiser \& Squires(1993)}]{kaiser_squires_1993}
Kaiser, N., \& Squires, G. 1993, Astrophys. J., 404, 441

\bibitem[{{Kocsis} {et~al.}(2006){Kocsis}, {Frei}, {Haiman}, \&
  {Menou}}]{kocsis_frei_etal_2006}
{Kocsis}, B., {Frei}, Z., {Haiman}, Z., \& {Menou}, K. 2006, \apj, 637, 27,
  arXiv:astro-ph/0505394

\bibitem[{{Kocsis} \& {Loeb}(2008)}]{kocsis_loeb_2008}
{Kocsis}, B., \& {Loeb}, A. 2008, \prl, 101, 041101, 0803.0003

\bibitem[{{Kopparapu} {et~al.}(2008){Kopparapu}, {Hanna}, {Kalogera},
  {O'Shaughnessy}, {Gonz{\'a}lez}, {Brady}, \&
  {Fairhurst}}]{kopparapu_hanna_etal_2008}
{Kopparapu}, R.~K., {Hanna}, C., {Kalogera}, V., {O'Shaughnessy}, R.,
  {Gonz{\'a}lez}, G., {Brady}, P.~R., \& {Fairhurst}, S. 2008, \apj, 675, 1459,
  0706.1283

\bibitem[{{Lang} \& {Hughes}(2006)}]{lang_hughes_2006}
{Lang}, R.~N., \& {Hughes}, S.~A. 2006, \prd, 74, 122001, arXiv:gr-qc/0608062

\bibitem[{{Laureijs} {et~al.}(2008)}]{euclid_science_2008}
{Laureijs}, R., {et~al.} 2008, ESA Science document (DEM-SA-Dc-00001),
  Available 6 July 2009 @
  http://sci.esa.int/science-e/www/object/index.cfm?fobjectid=42822

\bibitem[{{Massey} {et~al.}(2007){Massey}, {Rowe}, {Refregier}, {Bacon}, \&
  {Berg{\'e}}}]{massey_rowe_etal_2007}
{Massey}, R., {Rowe}, B., {Refregier}, A., {Bacon}, D.~J., \& {Berg{\'e}}, J.
  2007, \mnras, 380, 229, arXiv:astro-ph/0609795

\bibitem[{{NASA}(2009)}]{jdem_web}
{NASA}. 2009, JDEM Project Webpage, Available 12 June 2009,
  http://jdem.gsfc.nasa.gov/

\bibitem[{{Nissanke} {et~al.}(2009){Nissanke}, {Hughes}, {Holz}, {Dalal}, \&
  {Sievers}}]{nissanke_hughes_etal_2009}
{Nissanke}, S., {Hughes}, S.~A., {Holz}, D.~E., {Dalal}, N., \& {Sievers},
  J.~L. 2009, ArXiv e-prints, 0904.1017

\bibitem[{{Okura} {et~al.}(2007){Okura}, {Umetsu}, \&
  {Futamase}}]{okura_umetsu_etal_2007}
{Okura}, Y., {Umetsu}, K., \& {Futamase}, T. 2007, \apj, 660, 995,
  arXiv:astro-ph/0607288

\bibitem[{{Porter} \& {Cornish}(2008)}]{porter_cornish_2008}
{Porter}, E.~K., \& {Cornish}, N.~J. 2008, \prd, 78, 064005, 0804.0332

\bibitem[{{Refregier} {et~al.}(2008){Refregier}, {Amara}, {Kitching}, \&
  {Rassat}}]{icosmo_2008}
{Refregier}, A., {Amara}, A., {Kitching}, T., \& {Rassat}, A. 2008, ArXiv
  e-prints, 0810.1285

\bibitem[{{Rimes} \& {Hamilton}(2005)}]{rimes_hamilton_2005}
{Rimes}, C.~D., \& {Hamilton}, A.~J.~S. 2005, \mnras, 360, L82,
  arXiv:astro-ph/0502081

\bibitem[{{Schneider} \& {Er}(2008)}]{schneider_er_2008}
{Schneider}, P., \& {Er}, X. 2008, \aap, 485, 363, 0709.1003

\bibitem[{{Schutz}(1986)}]{schutz_1986}
{Schutz}, B.~F. 1986, \nat, 323, 310

\bibitem[{{Seitz} \& {Schneider}(1995)}]{seitz_schneider_1995}
{Seitz}, C., \& {Schneider}, P. 1995, \aap, 297, 287, arXiv:astro-ph/9408050

\bibitem[{{Sesana} {et~al.}(2007){Sesana}, {Volonteri}, \&
  {Haardt}}]{sesana_volonteri_etal_2007}
{Sesana}, A., {Volonteri}, M., \& {Haardt}, F. 2007, \mnras, 377, 1711,
  arXiv:astro-ph/0701556

\bibitem[{{Shapiro}(2009)}]{shapiro_2009}
{Shapiro}, C. 2009, \apj, 696, 775, 0812.0769

\bibitem[{{Smith} {et~al.}(2003){Smith}, {Peacock}, {Jenkins}, {White},
  {Frenk}, {Pearce}, {Thomas}, {Efstathiou}, \&
  {Couchman}}]{smith_peacock_etal_2003}
{Smith}, R.~E. {et~al.} 2003, \mnras, 341, 1311, arXiv:astro-ph/0207664

\bibitem[{{Vecchio}(2004)}]{vecchio_2004}
{Vecchio}, A. 2004, \prd, 70, 042001, arXiv:astro-ph/0304051

\bibitem[{{White}(2005)}]{white_2005}
{White}, M. 2005, Astroparticle Physics, 23, 349, arXiv:astro-ph/0502003

\end{thebibliography}
\bibliographystyle{hapj}
\end{document}